\documentclass[sigconf, nonacm]{acmart}
\AtBeginDocument{%
  }

\setcopyright{rightsretained}
\copyrightyear{2026}

\usepackage{listings}
\usepackage{xcolor}

\definecolor{codegreen}{rgb}{0,0.6,0}
\definecolor{codegray}{rgb}{0.5,0.5,0.5}
\definecolor{codepurple}{rgb}{0.58,0,0.82}
\definecolor{backcolour}{rgb}{0.95,0.95,0.92}

\lstdefinelanguage{json}{
    basicstyle=\ttfamily\footnotesize,
    numbers=left,
    numberstyle=\tiny\color{codegray},
    stepnumber=1,
    numbersep=8pt,
    showstringspaces=false,
    breaklines=true,
    frame=lines,
    stringstyle=\color{codepurple},
    literate=
     *{0}{{{\color{magenta}0}}}{1}
      {1}{{{\color{magenta}1}}}{1}
      {2}{{{\color{magenta}2}}}{1}
      {3}{{{\color{magenta}3}}}{1}
      {4}{{{\color{magenta}4}}}{1}
      {5}{{{\color{magenta}5}}}{1}
      {6}{{{\color{magenta}6}}}{1}
      {7}{{{\color{magenta}7}}}{1}
      {8}{{{\color{magenta}8}}}{1}
      {9}{{{\color{magenta}9}}}{1}
      {:}{{{\color{codegray}{:}}}}{1}
      {,}{{{\color{codegray}{,}}}}{1}
      {\{}{{{\color{codegray}{\{}}}}{1}
      {\}}{{{\color{codegray}{\}}}}}{1}
      {[}{{{\color{codegray}{[}}}}{1}
      {]}{{{\color{codegray}{]}}}}{1},
}

\lstdefinelanguage{markdown}{
    basicstyle=\ttfamily\footnotesize,
    numbers=left,
    numberstyle=\tiny\color{codegray},
    stepnumber=1,
    numbersep=8pt,
    showstringspaces=false,
    breaklines=true,
    frame=single,
    rulecolor=\color{codegray},
    commentstyle=\color{codegreen},
    keywordstyle=\color{blue},
    stringstyle=\color{codepurple},
    morekeywords={1., 2., 1}, 
}

\begin{document}

\newcommand{\singleAgentBaseline}{single\_agent\_baseline}

\title{A Multi-agent AI System for Deep Learning Model Migration from TensorFlow to JAX}

\author{Stoyan Nikolov}
\authornote{These authors contributed equally to this research.}
\email{stoyannk@google.com}
\affiliation{%
  \institution{Google}
  \city{Munich}
  \country{Germany}
}

\author{Bernhard Konrad}
\authornotemark[1]
\email{bernhardkonrad@google.com}
\affiliation{%
  \institution{Google}
  \city{Munich}
  \country{Germany}
}

\author{Moritz Gronbach}
\authornotemark[1]
\email{moritzg@google.com}
\affiliation{%
  \institution{Google}
  \city{Munich}
  \country{Germany}
}

\author{Niket Kumar}
\email{niketkb@google.com}
\affiliation{%
  \institution{Google}
  \city{Mountain View}
  \state{California}
  \country{USA}
}

\author{Ann Yan}
\email{annyan@google.com}
\affiliation{%
  \institution{Google}
  \city{Mountain View}
  \state{California}
  \country{USA}
}

\author{Varun Singh}
\email{varunsng@google.com}
\affiliation{%
  \institution{Google}
  \city{Mountain View}
  \state{California}
  \country{USA}
}

\author{Yaning Liang}
\email{yaning@google.com}
\affiliation{%
  \institution{Google}
  \city{Mountain View}
  \state{California}
  \country{USA}
}

\author{Parthasarathy Ranganathan}
\email{parthas@google.com}
\affiliation{%
  \institution{Google}
  \city{Sunnyvale}
  \state{California}
  \country{USA}
}

\renewcommand{\shortauthors}{Nikolov, Konrad, Gronbach, et al.}

\renewcommand{\shortauthors}{Nikolov, Konrad, Gronbach et al.}

\begin{abstract}
  The rapid development of AI-based products and their underlying models has led to constant innovation in deep learning frameworks. Google has been pioneering machine learning usage across dozens of products. Maintaining the multitude of model source codes in different ML frameworks and versions is a significant challenge. So far the maintenance and migration work was done largely manually by human experts.
We describe an AI-based multi-agent system that we built to support automatic migration of TensorFlow-based deep learning models into JAX-based ones. We make three main contributions: First, we show how an AI planner that uses a mix of static analysis with AI instructions can create migration plans for very complex code components that are reliably followed by the combination of an orchestrator and coders, using AI-generated example-based playbooks.
Second, we define quality metrics and AI-based judges that accelerate development when the code to evaluate has no tests and has to adhere to strict style and dependency requirements. Third, we demonstrate how the system accelerates code migrations in a large hyperscaler environment on commercial real-world use-cases.
Our approach dramatically reduces the time (6.4x-8x speedup) for deep learning model migrations and creates a virtuous circle where effectively AI supports its own development workflow.
We expect that the techniques and approaches described here can be generalized for other framework migrations and general code transformation tasks.

\end{abstract}





\maketitle

\section{Introduction}
As AI adoption is exploding, the quality and costs of operating the ML models that power this revolution are becoming critical to success. The key building blocks of these models are the ML frameworks they are written in. Google builds and operates thousands of machine learning models across its product suite and many are built on top of TensorFlow  \cite{abadi2016tensorflow}. The company is now shifting to JAX \cite{jax2018github} as a standard framework for the development of ML models which is better suited for Tensor Processing Unit (TPU) \cite{jouppi2023tpu} inference and training, and underpins the vast majority of the Google AI stack.
Although most new development within Google is happening on JAX, there still are thousands of models that need to be migrated from TensorFlow - the other standard framework for ML development that Google created. Manually migrating all of them would entail hundreds of expert engineering years as the two ML frameworks are not equivalent feature-wise and have different architectures and paradigms. The effort would be better spent on actually improving the architectures of the models.

AI-powered tools like Gemini CLI \cite{gemini2025cli}, Codex CLI \cite{openai2026codexcli}, Claude Code \cite{anthropic2025claudecode}, etc. have significantly simplified the workflows of software engineers and can tackle increasingly hard problems. However, for very complex code migrations over many components, where a lot of specific context is needed, it is still challenging to produce reliable and high quality output \cite{stackoverflow2025survey}. We assert that a bespoke system focused on a class of problems can significantly simplify and shorten the development time. 

To solve the problem we built a system for migrations focused on TensorFlow to JAX Flax Linen \cite{flax2020github} which has shown strong positive results and allows landing reworked models in a fraction of the time compared to manual migration while addressing quality challenges of more general AI agentic systems.

This article is a detailed case-study of the solution we built, how we arrived at it, the metrics we chose to evaluate it, and the results we achieved. We believe that our solution demonstrates a novel approach to code migrations for machine learning models and that the presented techniques are applicable in other code transformation domains as well.

An interesting aspect of our system is that we effectively use AI techniques to improve AI development. We feel that this virtuous improvement cycle is going to persist in the code migration space and elsewhere in the industry, and can further accelerate the progress and adoption of AI.

The paper is organized as follows. Section~\ref{section:background} provides the background and problem statement, detailing the architectural differences between TensorFlow and JAX. In Section~\ref{section:design}, we present our multi-agent design, including the hierarchy of playbooks and the specialized roles of the planner, orchestrator, and coder agents. Section~\ref{section:evaluation} outlines our evaluation methodology, describing our use of open-source and internal models alongside a checklist-based LLM judge for quality assessment. In Section~\ref{section:results}, we present the results of our experiments and ablation studies, highlighting significant time savings and quality improvements. Section~\ref{section:threats} discusses threats to validity, Section~\ref{section:related} reviews related work, and Section~\ref{section:conclusion} provides concluding remarks.

\section{Background and problem statement} \label{section:background}
TensorFlow and JAX are both machine learning frameworks but are driven by differing motivations and eras. Developed in the pre-transformer era, TensorFlow was designed around the predominant methodologies of the time, such as CNNs operating on standard GPUs. It relies on stateful layers and static execution graphs, often utilizing specialized custom operations for performance. In contrast, JAX represents a shift towards dynamic execution and is built on a functional, stateless paradigm optimized for modern TPU infrastructure and XLA compilation \cite{jax2018github}.

These architectural divergences create significant migration challenges. TensorFlow's object-oriented, stateful layer initialization—often abstracted through the Keras API—contrasts sharply with JAX's functional approach. In JAX, state is explicitly managed and separated from model definition, requiring a fundamental rethinking of how layers interact. Furthermore, many production TensorFlow models rely on specialized "custom ops" that represent near-hardware-level code. These rarely have direct equivalents in JAX and often require reimplementation from first principles using JAX's composable primitives. Examples of some of the differences between the frameworks are visible in Listings \ref{lst:tf_model} and \ref{lst:flax_model}.

Beyond the core model code, production migrations must also address the surrounding ecosystem. This includes execution scripts, data loading pipelines, and training loops, all of which are deeply coupled with the framework. Additionally, migrated code must adhere to strict internal coding standards and reuse existing libraries. Naive application of generic coding agents often yields unsatisfactory results, failing to capture complex metrics or respect specific style guides. Consequently, effective migration requires a system capable of handling deep domain context and maintaining functional parity across thousands of lines of code.

\begin{figure}[ht]
\centering
\begin{minipage}{\linewidth}
\begin{lstlisting}[caption={Standard TensorFlow/Keras implementation with implicit state management and regularization side-channels.}, label={lst:tf_model}]
import tensorflow as tf

class TFModel(tf.keras.Model):
  def __init__(self, units=64):
    super().__init__()

    # 1. BatchNorm: State updated automatically via 'training' flag.
    # Defaults: momentum=0.99, epsilon=0.001
    self.bn = tf.keras.layers.BatchNormalization()

    # 2. Hyperparameters default: alpha=0.3
    self.act = tf.keras.layers.LeakyReLU() 
 
    # 3. Regularization: Implicitly tracked via side-channel (model.losses)
    self.dense = tf.keras.layers.Dense(
      units, kernel_regularizer=tf.keras.regularizers.l2(1e-4)
    )

  def call(self, x, training=False):
    x = self.dense(x)
    x = self.bn(x, training=training)
    return self.act(x)
\end{lstlisting}
\end{minipage}
\end{figure}

\begin{figure}[ht]
\centering
\begin{minipage}{\linewidth}
\begin{lstlisting}[caption={Migrated Flax implementation requiring explicit state handling for BatchNormalization and manual PyTree transformations for L2 regularization.}, label={lst:flax_model}]
import jax.numpy as jnp
import flax.linen as nn

class FlaxModel(nn.Module):
  units: int = 64

  @nn.compact
  def __call__(self, x, train: bool):
    x = nn.Dense(features=self.units)(x)
    
    # 1a. BatchNorm: Flax default: epsilon=1e-5. 
    # Must override to match TF's numerical behavior.
    x = nn.BatchNorm(
        use_running_average=not train, 
        momentum=0.99, 
        epsilon=0.001
    )(x)
    
    # 2. Hyperparameters: Flax default: negative_slope=0.01.
    # Must override to match TF's default alpha=0.3.
    return nn.leaky_relu(x, negative_slope=0.3)

def loss_fn(params, batch_stats, x, y):
  # 1b. State Management: Must explicitly pass and retrieve mutated batch_stats
  logits, mutated_vars = model.apply(
      {'params': params, 'batch_stats': batch_stats},
      x, train=True, mutable=['batch_stats']
  )
  
  # 3. Regularization: Manually transform the 'params' PyTree to calculate loss.
  # This replaces the 'kernel_regularizer' side-channel in TF.
  l2_loss = 0.5 * sum(jnp.sum(jnp.square(p)) for p in jax.tree_util.tree_leaves(params))
  
  main_loss = optax.softmax_cross_entropy(logits, y).mean()
  return main_loss + 1e-4 * l2_loss, mutated_vars['batch_stats']
\end{lstlisting}
\end{minipage}
\end{figure}

\section{Our Design} \label{section:design}

Google has been investing for the last few years in an AI-based system for code migrations as described in \cite{nikolov2025googleusingaiinternal}. The system is very flexible and we extended and adapted it to support the TensorFlow to JAX migration with a multi-agent architecture.

\subsection{Prompt Engineering}

To imbue the needed knowledge into the agents about the structure and style of the expected code, we built a hierarchy of playbooks - markdown files provided to the agent.

\subsubsection{Playbook generation and usage}

The initial development phase focused on testing the AI agent's capability for migrating open source models from TensorFlow to JAX. This revealed three primary challenges:

\begin{itemize}
    \item API hallucinations
    \item Style inconsistency
    \item Inability to produce consistently buildable code
\end{itemize}

To resolve these issues, we evolved from basic prompt instructions to a playbook system, which provides targeted guidance to the AI agent.

\begin{table*}[t]
  \caption{The different playbooks, and how they are organized based on their content, role in execution, and update cadence.}
  \label{tab:playbooks}
  \centering
  \small
  \begin{tabular}{lp{5cm}p{4cm}p{4cm}}
    \toprule
    \textbf{Playbook Type} & \textbf{Focus and Content} & \textbf{Role in Execution} & \textbf{Update Cadence} \\
    \midrule
    General Instruction & Provides fundamental capabilities, instructing the agent on tools necessary to read, write, search, and interact with Google's monorepo system. & Used by almost every agent to interact with Google's monorepo. & Updated only when new tools are added. \\
    \addlinespace
    Style Playbook & Developed for focused tasks, such as a style playbook containing Google-specific coding style rules. & Used specifically for coding tasks. & Updated very rarely and manually when style guide violations are detected. \\
    \addlinespace
    Task-Specific & Tailored for specific tasks, such as the TensorFlow to Flax Linen migration. & Provides generally applicable instructions for coding tasks. & Updated whenever we discover new generally applicable instructions for the task. \\
    \addlinespace
    Client-Specific (e.g., YouTube) & Contains use-case-specific requirements and preferred libraries. & The primary source of task-relevant information, used as part of the prompt for both planning and coding tasks. & Updated iteratively after each full execution to improve capabilities for the specific task (e.g., refining imports). \\
    \bottomrule
  \end{tabular}
\end{table*}

The playbooks are in standard markdown format, which is familiar to modern LLMs and Gemini in particular. The playbooks are concatenated and provided as part of the system prompt to the agent (usually the coder agent), although we internally split them into the hierarchical set shown in Table~\ref{tab:playbooks}.

The **General Instruction Playbook** covers repository specifics like directory structures and build tools, shared across most internal coding agents. The **Style Playbook** enforces consistency in areas with multiple valid options (e.g., using `nn.compact` over `setup()`, consistent imports). The **Task-Specific Playbook** was generated by Gemini from official documentation to highlight subtle API differences (e.g., default parameters) between TensorFlow and JAX. Finally, **Client-Specific Playbooks** leverage few-shot examples \cite{brown2020languagemodelsfewshotlearners} to capture team-specific conventions and patterns.

\subsubsection{Strategy for Client-Specific Playbook Generation}

When completed migrations are available, we leverage these as "golden examples" to generate a client-specific playbook offline. To ensure scalability and avoid overfitting to these initial migrated models, the following methodology was adopted:

\begin{itemize}
\item We prompt an AI agent to break down the migrated code into semantic units (e.g., base imports, base methods, loss computation, metrics output).
\item An LLM then summarizes general migration rules based on these code blocks.
\end{itemize}

The generation was done by Gemini but supervised by a human engineer. The human engineer is mostly responsible for reviewing the issues originating due to TensorFlow and JAX framework differences. The LLM prompt breaks down the two code bases into semantic domain (machine learning) code blocks, which are functionally equivalent. The broken down code blocks tend to be fine grained, which helps to generate playbooks distilled with both some generic and client-specific knowledge base.

This process created a client-specific playbook that contained both sufficient information and practical coding examples, proving scalable in handling new, unseen models well (see Section~\ref{section:evaluation}). In our experiments we were able to generate a good enough playbook from just 2 examples - one we deemed mid-complexity and one highly complex model with thousands of lines of code and dozens of layers.

For situations where no already-migrated models and examples exist we would rely only on the generic instructions and expect that the humans would need to further improve the quality of the migrated code. In the Results section \ref{section:results} we see the quality improvement coming directly from an auto-generated YouTube playbook.

As more complex models get migrated, we can repeat the above process to improve the playbook, creating a virtuous cycle.

\subsection{A Multi-agent System}

\subsubsection{Motivation}

Single-agent setups struggled with long-horizon tasks, frequently losing context or hallucinating completion on complex models. A multi-agent decomposition was necessary to enforce verification and manage context windows. This approach addresses the complexity of tasks that require following strict development rules, handling deep dependencies, and ensuring migration completeness for models spanning multiple files.

\begin{figure*}[t]
  \centering
  \includegraphics[width=\textwidth]{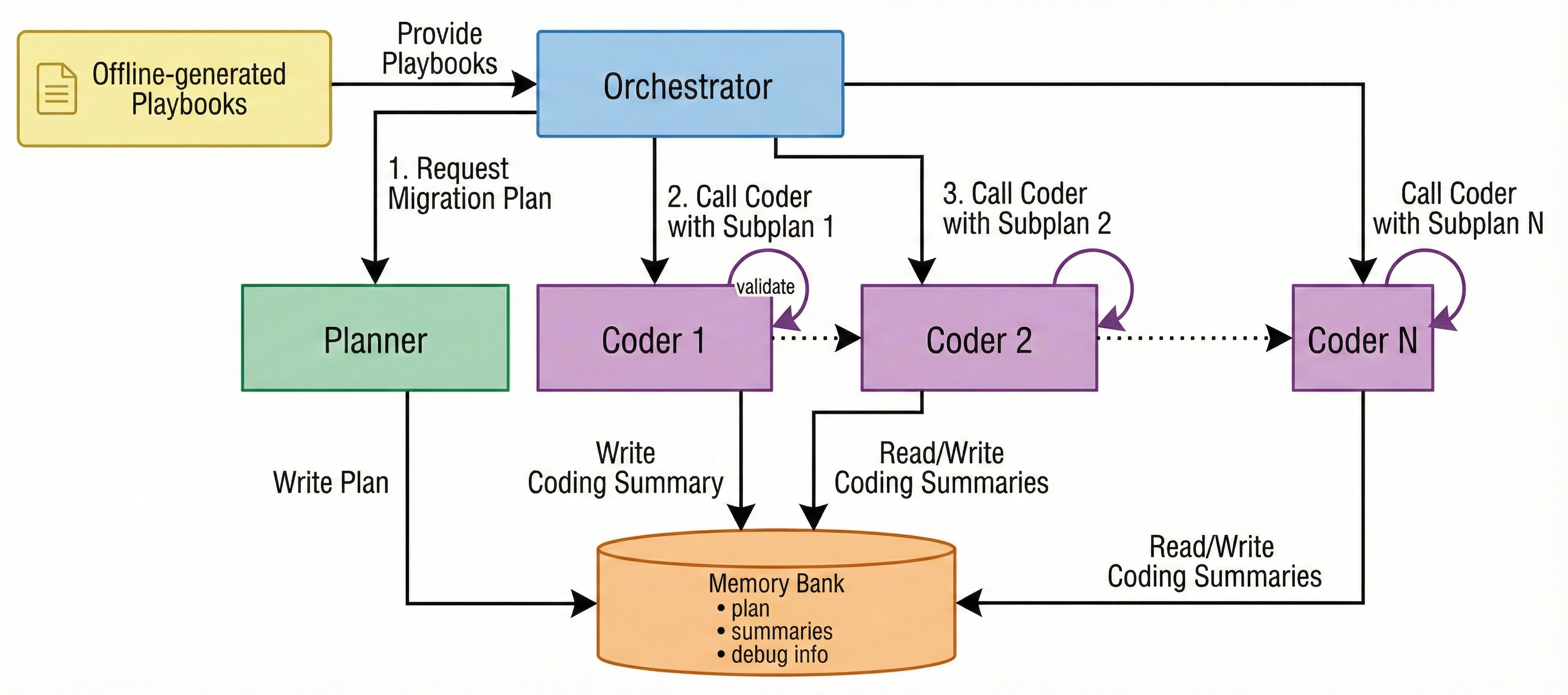}
  \caption{System diagram}
  \label{fig:diagram}
  \Description{System diagram}
\end{figure*}

We introduced two additional agents: An orchestrator and a planner. Neither writes code, but they guide the coder through the task and are critical in keeping the scope of each invocation of the coder manageable. The planner investigates and summarizes dependencies, which is then translated into a migration plan. The orchestrator splits the full plan into smaller, manageable and verifiable chunks. Each chunk is passed to an independent coder, sequentially. The full system diagram can be seen in Figure \ref{fig:diagram}.

\begin{itemize}
    \item \textbf{Planner}
    \begin{itemize}
        \item Creates an end-to-end plan with disjoint steps.
        \item A coding task may be split into multiple steps.
        \item Leans towards fine-grained steps, requiring not more than a few hundred new LOC each.
        \item Adds a validation condition that must be met for a step to be complete.
    \end{itemize}
    
    \item \textbf{Orchestrator}
    \begin{itemize}
        \item Combines one or more logically connected steps from the generated plan into coarse ``sub-steps'' based on configurable strategies.
        \item Delegates one sub-step at a time to the coder.
        \item Waits for the response from the coder before providing the next sub-step.
    \end{itemize}
    
    \item \textbf{Coder}
    \begin{itemize}
        \item Works on one ``sub-step'' at a time and responds back to the Orchestrator.
        \item Verifies the validation condition.
    \end{itemize}
\end{itemize}

Having multiple collaborating agents also allows for independent development and improvements of each agent. 

\subsubsection{Information-sharing between the agents}

To enable communication and state persistence without requiring direct agent-to-agent handoffs, we implemented a file-based memory bank. This shared storage serves as the system's "connective tissue", allowing agents to access shared context independently. It prevents drift and redundant work by storing migration state, debugging artifacts, and versioned snapshots of the playbooks used at each step.

\subsubsection{The Planner Agent}

Key to splitting the work into manageable steps is creating a plan for the migration of a single TensorFlow model. In our definition, a ‘step’ is a discrete instruction given to a coder agent which will produce code in a valid state - that is code that builds and, if applicable, is testable. For a complex layer the planner will produce multiple, occasionally dozens of steps.

Splitting the work this way allows us to:
\begin{itemize}
    \item Control the scope of each coder agent's assignment.
    \begin{itemize}
        \item Guide the model across multiple related files.
        \item Keeps the task complexity for any coder agent small.
    \end{itemize}
    \item Increase the reproducibility of the changes.
    \item Allow manual inspection to validate that we have covered all needed elements to migrate.
\end{itemize}

\paragraph{Step Granularity}

The step granularity varies, but examples include instructing the model to create the desired (empty) folder structure, migrating specific metrics, functions, classes.
Our TensorFlow models usually span multiple files, including distant folders for dependencies. In order to reduce the complexity for the coding tasks, the responsibility of deeply understanding the file and component hierarchy of the model is given to the planner. The steps that the planner will generate will specify which files need to be opened and edited by the coder, freeing the coder from having to search the file system independently.
We experimented with multiple techniques to create the planner, but in the end opted for a hybrid static-agentic approach. Given that we know only the main model file that has to be migrated, we need to discover all its dependencies and eventually which ones of them have to also be migrated or might already have counterparts in JAX. The actual migration has to happen upwards from the ‘leaf’ nodes - layers or functions without non-migrated dependencies. The goal is that every step of the plan is buildable and ideally testable.

\paragraph{Dependency Discovery}

For dependency discovery we don’t use an agent, but instead rely on Kythe \cite{google2025kythe}, a tool for compiler-based static analysis and cross-references. Kythe resolves dependencies based on a recent index of the codebase, enhanced by build information. This makes dependency discovery deterministic, verified, fast, and computationally cheap.

The planning is done recursively. Starting with just the main model file and its direct dependencies, we prompt an LLM to come up with a migration plan, identify gaps in the plan, and to specify a list of files where dependencies are required to fill those gaps. We repeat this process, adding the content of all requested dependencies to the prompt.
General JAX libraries that are ready for use are injected into the planner prompt (via the playbooks), to avoid creating a migration step for those.

\paragraph{The Plan Schema}

Each produced migration step includes an index, original and target file paths, instructions for the coder, and indices to dependent steps.

The plan is provided in JSON format (see Listing \ref{lst:plan_schema}) as we discovered it suitable for the LLM to follow, while still being easy for humans to read.

\begin{figure}[ht]
\centering
\begin{minipage}{\linewidth}
\begin{lstlisting}[language=json, caption={Example migration plan schema generated by the Planner Agent, defining step-level instructions and cross-step dependencies.}, label={lst:plan_schema}]
{
  "steps": [
    {
      "step_id": 1,
      "title": "Create JAX Model Directory and BUILD File",
      "target_files": [ "//.../jax/BUILD" ],
      "instructions": "Create a new directory named `jax`... create an empty file named `BUILD`.",
      "validation": "The file `//.../jax/BUILD` exists and is empty."
    },
    "...",
    {
      "step_id": 7,
      "title": "Migrate `FeatureSelection` Layer to `feature_select.py`",
      "source_files": [ "//.../tf/feature_select.py" ],
      "target_files": [ "//.../jax/feature_select.py" ],
      "instructions": "Create a new file... Convert it from a `tf.keras.layers.Layer` to a `flax.linen.Module`...",
      "validation": "The file ... contains a `FeatureSelection` Flax Module...",
      "dependencies": [ 5 ]
    }
  ]
}
\end{lstlisting}
\end{minipage}
\end{figure}

\paragraph{Interactions with the Memory Bank}

The planner writes its complete plan to the memory bank, in JSON format. For visualization and quick human inspection, we also produce a dot version of the plan.

\subsubsection{The Orchestrator Agent}

The orchestrator serves as the bridge between the high-level plan and the Coder's execution, managing the migration process without modifying code directly. To optimize LLM performance, it employs \textbf{dynamic task scoping}, grouping plan steps into manageable chunks based on complexity to keep the Coder's context focused. It also performs \textbf{dynamic knowledge injection}, selecting only relevant playbooks for each task to minimize context pollution. Additionally, the orchestrator provides \textbf{strategic failure handling}—deciding to retry, skip, or abort based on Coder outcomes—and ensures \textbf{fault tolerance} by persisting progress to the memory bank, allowing long-running migrations to resume after interruptions.

\subsubsection{The Coder Agent}

The coding agent is a ReAct agent [\cite{yao2023react}] with access to basic tools to list files, read files, code search (grep-like for the Google3 repository), write files, search and replace, auto-fix dependencies, run builds, and run unit tests.

The system prompt contains our general playbook on how to work within Google’s repository, and specifies a workflow that includes building and testing each change. It also contains the additional use-case specific instructions, such as for YouTube models.

We use the coding agent as the “workhorse” of the migration. It is supposed to generate validated compilable and testable code each time it is invoked.

A critical capability of the agent is to build and test the code it creates. This also implies each prompt it receives (one or multiple steps from the plan) should lead to logically complete code that can be built. The coding agent is instructed to fix any failures in the builds and tests it encounters and to continue the process until it has completed its given assignment (a plan step).

The main final artifacts of the coding agent are the file changes that implement the given assignment. 

\paragraph{Validation}

To prevent premature completion—a common failure mode in long-context generation—we implemented a self-correction step where the agent reviews its own output against the task requirements before finalizing. Once the coder agent declares it is finished, we artificially insert an additional prompt asking it to validate that the task has been completed successfully. This simple "self-review" effectively eliminates missed steps without the overhead of a separate reviewer agent.

\paragraph{Interactions with the Memory Bank}

A common failure mode in multi-agent systems is the loss of project-level context as the migration progresses.  To mitigate this, the coder creates and stores a summary (see Listing \ref{lst:coder_summary}) of the changes made. This helps the orchestrator understand the state of the changes and forms a ‘short-term memory’ across the migration task. When invoking the Coder, we always pass along the full history of created summaries as part of the initial prompt.
\begin{figure}[ht]
\centering
\begin{minipage}{\linewidth}
\begin{lstlisting}[language=markdown, caption={Example of a Markdown summary generated by the Coder Agent after a step execution, serving as project-level context for subsequent tasks.}, label={lst:coder_summary}]
Coder summary of Step N
1. Changes Made
- Created //.../model_utils.py: Implemented weighting functions, loss helpers, and custom JAX metrics (MSE, AUCROC).
- Created //.../model_utils_test.py: Unit tests for all utility functions.
- Updated //.../BUILD: Added new targets and dependencies.

2. Key Fixes & Learnings
- Dependency Resolution: Fixed ModuleNotFoundError for clu and metrax by adding explicit deps to the BUILD file.
\end{lstlisting}
\end{minipage}
\end{figure}

By presenting the coder with a list of absolute paths and the high-level logic changed in previous steps, the agent gains a global overview of the migration. This ensures it knows exactly where the new JAX components live and prevents it from inadvertently undoing previous work or re-implementing existing logic.
In the rare case when the coding agent is not able to produce code that builds, it will communicate this back to the orchestrator, which will decide how to proceed.

\section{Evaluation} \label{section:evaluation}

We built two evaluation sets with golden examples. Having objective quality benchmarks increased our development speed as it allowed for easy verification of the impact of new features of the agentic system.

\subsection{Open Source Models}

Our initial evaluation utilized a set of 32 open source models, which was instrumental in refining the system and eliminating early tool and model definition issues. However, with the introduction of Gemini 3 Pro \cite{deepmind2026gemini}, the performance on this dataset saturated, shifting its role to primarily regression testing.

\subsection{Google-internal models}

\subsubsection{Playbook-Generation Set}

We built a small initial evaluation suite based on two real YouTube Google models. These had both still active legacy TensorFlow versions and a human-migrated JAX version.
The goal of this evaluation set was to test our solution on real models which must strictly follow internal style guides and reuse internal components. One of the models we deemed to be of moderate complexity, while the second was extremely complex, with thousands of lines of JAX code and dozens of internal metrics. Especially the complexity of the second model pushed us to implement a detailed planner and semi-static plan building to help the agent keep track of all the functions, classes and metrics it would have to migrate.
We used these models also to generate our playbooks which carries a risk of overfitting the prompts. We also took the practical approach that in real world scenarios, we would only encounter a handful of human-migrated models and would need to use them to generalize for the dozens or hundreds more.
We demonstrate in the Evaluation section \ref{section:evaluation} that indeed the generated prompts generalize to other YouTube models.

\subsubsection{Test Set}

We evaluate six moderate-to-high complexity real-world YouTube models (see Table \ref{tab:model_complexity}). This excludes the two models from the training set that we iterated on for prompt engineering. We assume a high complexity model has > 2000 lines of code, more than 100 layers, dependencies across modules and more than 100 metrics.
In this case the complexity is purely in terms of ‘code complexity’ and dictated by where we saw the simple coding agents struggle.

\begin{table}[ht]
  \caption{Complexity metrics for the YouTube models in the evaluation test set.}
  \label{tab:model_complexity}
  \centering
  \small
  \begin{tabular}{lccc}
    \toprule
    \textbf{Model} & \textbf{Line Count (Main/Deps.)} & \textbf{Layers} & \textbf{Metrics} \\
    \midrule
    small\_1 & 0.5k / $\sim$500 & $\sim$20 & $\sim$20 \\
    small\_2 & 0.4k / $\sim$500 & $\sim$17 & $\sim$25 \\
    medium\_1 & 2.2k / $\sim$1k & $\sim$40 & $\sim$30 \\
    medium\_2 & 2.3k / $\sim$3k & $\sim$40 & $\sim$60 \\
    large\_1 & 2.7k / $\sim$13k & $\sim$250 & $\sim$400 \\
    large\_2 & 2.9k / $\sim$9k & $\sim$130 & $\sim$210 \\    
    \bottomrule
  \end{tabular}
\end{table}

For each of these six models we created a checklist as described in subsection \ref{subsection:quality_metrics}, which was used to score the JAX migration.

\subsection{Quality Metrics} \label{subsection:quality_metrics}

Standard code metrics like edit distance or AST comparison fail to capture the functional correctness of complex ML logic (e.g., gradient scaling, loss masking). To address this, we developed an evaluation strategy centered on human-defined criteria and automated "blind" audits, allowing us to assess quality improvements and identify migration faults even without a perfect ground truth.

\subsubsection{Checklist-based LLM Judge}

The primary metric for our system is a granular, framework-agnostic checklist. For each model in our evaluation set, we manually curate a "Golden Checklist" that captures the essential elements the migrated model must possess.
These checklists are intentionally designed to be framework-agnostic (but not use-case agnostic). Instead of referencing TensorFlow-specific syntax, they use generic layer names or descriptive requirements (e.g., "Towers for each prediction head" or "User embeddings are L2 normalized", see Listing \ref{lst:model_checklist}). This ensures the checklist remains a stable target even as the underlying implementation changes.
The evaluation is performed by a dedicated "Judge" LLM-based agent, again powered by Gemini 3 Pro. The process follows a blind audit pattern:

\begin{enumerate}
    \item \textbf{Input:} The Judge is provided with the checklist and the location of the migrated JAX model files.
    \item \textbf{Inspection:} The agent is allowed to explore the code base from the main model file outwards to verify the implementation.
    \item \textbf{Scoring:} The Judge does not have access to the original TensorFlow source code. It must verify the checklist items based solely on the JAX code provided. This prevents the agent from simply comparing syntax (and effectively doing the same as the Coder) and forces it to validate the functional logic described in the checklist.
    \item \textbf{Results:} Each item is scored individually. The final quality score is the percentage of satisfied items.
\end{enumerate}

 The checklists primarily measure completeness of the migration. A perfect checklist score does not guarantee correctness. This approach is a deliberate compromise. While it lacks the absolute certainty of numerical parity, it is significantly faster to implement and provides a clear signal on whether the key parts of the model were captured. To minimize the impact of LLM scoring variance, we aim for checklist items that are as explicit and objective as possible.
The final score is the number of checked items divided by the total number of items in the checklist.

This metric is strictly used for post-migration scoring and is not fed back into the agents for automated fixes. We made this choice to maintain the integrity of the evaluation; using the metrics for an automated "fix-it loop" would make them less effective for comparing different architectural approaches or model versions.

\begin{figure}[ht]
\centering
\begin{minipage}{\linewidth}
\begin{lstlisting}[language=markdown, caption={Example of a framework-agnostic checklist used by the LLM Judge to evaluate migration completeness and architectural fidelity.}, label={lst:model_checklist}]
# [REDACTED] Model Checklist

## Model Architecture
- [ ] Shared user/context embedding tower
- [ ] Towers for each prediction head
- [ ] Point-wise Attention layer

## Prediction Heads & Tasks
- [ ] [REDACTED First] prediction task
- [ ] [REDACTED Second] prediction task
- [ ] [REDACTED Third] prediction task
- [ ] [REDACTED Fourth] prediction task

## Metrics
- [ ] [REDACTED] for classification tasks
- [ ] [REDACTED] for regression tasks
- [ ] Task-specific loss metrics
- [ ] [REDACTED] for casual users

## Predicted Outputs
- [ ] [REDACTED]
- [ ] [REDACTED]

## Correctness & Logic
- [ ] Input features are correctly processed
- [ ] User embeddings are L2 normalized
- [ ] ReLU activation in hidden layers
- [ ] Cross-Entropy loss for classification
- [ ] Downsampling frequent users
- [ ] Gradient scaling or stopping applied
- [ ] Bias correction is applied if needed
- [ ] Multi-task loss weighting is in place
\end{lstlisting}
\end{minipage}
\end{figure}
After we achieved a high enough score for the migrated models we did human-driven reviews to make sure that the expected quality was in place and that our metric is a stable proxy for real-world quality of the migrated models.

\subsubsection{Alternative Metrics Considered: Numerical Equivalence Tests}

By creating tests that execute both models on the same input with the same weights, we can get a high-fidelity score about migration success. This is the approach we followed for the original evaluation set based on OSS models.
However, for the real-world models, we found that relying on automated numerical parity as a primary metric is difficult for several reasons:

\begin{itemize}
    \item \textbf{Interface Discrepancies:} Pre-generated tests to score migrations heavily depend on the public interface of the migrated model (class, attribute, and method names, layer names, signatures, order in which methods are called, \dots). Specifying this interface fully was difficult and time-consuming for the open-source models (in fact, we didn't complete this process once it became apparent that the OSS set is saturated).
    \item \textbf{Engineering Complexity:} Setting up these tests requires managing weight transfers and complex state initialization between frameworks, which can often be as difficult as the migration itself.
    \item \textbf{Expected Divergence:} Subtle differences in how JAX and TensorFlow handle operations (such as XLA kernel fusion or specific activation functions like sigmoid) mean that minor numerical shifts do not always indicate a ``failed'' migration.
\end{itemize}

Overall, pre-generated equivalency tests require high effort to correctly set up, making it difficult to scale it as an evaluation metric.
Rather than using it as a high-level quality metric, we treat it as an element of the coding task. The coder agents are instructed to generate and run their own equivalency tests as part of the migration steps. Even if this doesn’t result in a fully successful equivalence test due to the aforementioned reasons, it’s often helpful to fix other issues (such as type or shape mismatches).

\section{Results} \label{section:results}

In this section we demonstrate the quality results we achieved for our multi-agent system as well as a comparison of the contributions of the key elements.

\subsection{Agent Configurations}

We are comparing the configuration listed in Table \ref{tab:agent_playbooks_reduced}:

\paragraph{single\_agent\_baseline:} A single autonomous ReAct coding agent with no specific adaptations for TensorFlow to JAX migrations, no planner, and no orchestrator. The agent can freely explore the codebase though and consult internal documentation.

\paragraph{single\_agent\_yt\_specific:} Same as \textit{single\_agent\_baseline}, but includes all TensorFlow to JAX playbooks (style, task-specific, \& client-specific). No planner and no orchestrator. This is the most sophisticated single agent.

\paragraph{multi\_agent:} A multi-agent system with planning \& orchestration with the Task-specific playbook, but without the client-specific playbook.

\paragraph{multi\_agent\_yt\_specific:} Same as \textit{multi\_agent}, but includes the client-specific playbook. This is the fully-featured configuration.

\begin{table}[ht]
    \caption{Agent Configuration Comparison}
    \label{tab:agent_playbooks_reduced}
    \centering
    \begin{tabular}{@{} l c c c @{}}
        \toprule
        \textbf{Configuration} & \textbf{Style \& Task} & \textbf{Client} & \textbf{Planner \&} \\
                                  & \textbf{Playbook}   & \textbf{Playbook} & \textbf{Orchestrator} \\
        \midrule
        single\_agent\_baseline      & No  & No  & No  \\
        single\_agent\_yt\_specific  & Yes & Yes & No  \\
        multi\_agent                 & Yes & No  & Yes \\
        multi\_agent\_yt\_specific   & Yes & Yes & Yes \\
        \bottomrule
    \end{tabular}
\end{table}

All agents are given the General Instructions Playbook and use Gemini 3 Pro with a temperature of 0.2 to induce slight variance.

\subsection{Migration Quality}

We run each configuration on all six YouTube models in the test set multiple times (N>10 each), each resulting in a migrated JAX model. Then, we run a Gemini judge that scores the migrated JAX model using the corresponding checklist. The checklist score is the completion ratio of the tasks on the checklist. For each model and agent configuration we then average the checklist scores. The total score per configuration is the average score over all six models.

Table~\ref{tab:agent_performance} shows the checklist scores of each agent configuration across all runs, all the migrated code \textbf{builds}.

\begin{table*}[t]
\caption{Performance Metrics Across Agent Configurations (Best Values Bolded)}
\label{tab:agent_performance}
\centering
\begin{tabular}{lcccc}
\toprule
\textbf{Model} & \textbf{multi\_agent} & \textbf{multi\_agent\_yt\_specific} & \textbf{single\_agent\_baseline} & \textbf{single\_agent\_yt\_specific} \\
\midrule
large\_1  & 0.31 & \textbf{0.65} & 0.09 & 0.14 \\
large\_2  & 0.42 & \textbf{0.59} & 0.22 & 0.33 \\
medium\_1 & 0.73 & \textbf{0.90} & 0.43 & 0.43 \\
medium\_2 & 0.53 & \textbf{0.75} & 0.38 & 0.44 \\
small\_1  & 0.70 & \textbf{0.95} & 0.66 & 0.48 \\
small\_2  & \textbf{0.87} & 0.82 & 0.59 & 0.36 \\
\bottomrule
\end{tabular}
\end{table*}

The total scores show that \textit{multi\_agent\_yt\_specific} is the strongest configuration (significantly, by paired t-test, $p < 0.02$), followed by the other multi-agent \textit{multi\_agent}. Then there is a significant gap ($p < 0.02$) to the single agents \textit{single\_agent\_yt\_specific} and \singleAgentBaseline{}, among which there is no statistical difference ($p = 0.62$). 

To understand common failure modes, we compared checklists for \singleAgentBaseline{} and \textit{multi\_agent\_yt\_specific}. The \textit{single\_agent\_baseline} agent usually gets the architecture of the model right, but frequently omits loss functions, evaluation metrics, and key logic such as label weighting.

In terms of computational cost, the multi-agent configurations take approximately 3-5 times longer than the single-agent configurations.

\subsection{Dependency on Model Complexity}

On the two simplest models, \textit{small\_1} and \textit{small\_2}, the significant difference between the \textit{single\_agent\_baseline} and \textit{multi\_agent\_yt\_specific} configurations vanished ($p = 0.07$). Generally, the more complex the TensorFlow model, the bigger the gap between the baseline and our main agent becomes. This result also aligns with our earlier finding that any of the four agent configurations, using Gemini 3 Pro, perform roughly equally well on the Open Source evaluation set, which consists of mostly small TensorFlow models.

As migrating more complex models is also more challenging for human developers, our results highlight the value of the multi-agent configuration for AI-based TensorFlow to JAX migrations of medium-to-large in-production models at scale.

\subsection{Time Savings}

We asked human experts to estimate the time taken for the models’ migrations for the small open source models - both manually and with our system. For these small models the migration speed was increased by a factor of 1.7x, and was mainly bound by the relatively short amount of manual work, and the fixed cost of reading the AI-generated code. However, the migration speed up increases significantly when the models grow in complexity and line count.

For two manually-migrated in-production models we gathered the time that the human expert required for the migration, and compared it to the time that a human expert required to get the code created by our main agent to a trainable model of equal quality. This required review and fixing of remaining issues.

\begin{table*}[t]
\caption{Migration Efficiency and Human-in-the-Loop Metrics}
\label{tab:speedup}
\centering
\small 
\begin{tabular}{lcccccr}
\toprule
\textbf{Model} & \textbf{Line Count} & \textbf{Layers} & \textbf{Metrics} & \textbf{Pure Manual Migration} & \textbf{AI plus Review \& Fix} & \textbf{Speed-up} \\
\midrule
model\_1 & 0.6k / 3k & 15  & 50  & $\sim$16 hrs & $\sim$2 hrs & 8x   \\
model\_2 & 4k / 20k  & 250 & 300 & $\sim$32 hrs & $\sim$5 hrs & 6.4x \\
\bottomrule
\end{tabular}
\end{table*}

We estimate time savings of 6.4x-8x (see Table~\ref{tab:speedup}) compared to human-only migration, signaling a dramatic improvement at scale.
The migration speed up is largest at medium-sized models, where human experts migration is complex due to the sophisticated model architecture, and the quality of the agent output is still high. As the models get more complex, the agent output also requires more human expert work to get it to the level of a human expert migration.

It should be noted that for the models in Table \ref{tab:speedup}, human experts did not write unit tests. Our migration system always does. In addition, the expertise required to review and fix a migrated model is arguably lower than the expertise required to migrate a model from scratch.

\section{Discussion} \label{section:discussion}

Our results show that \textit{multi\_agent\_yt\_specific} --- \textit{i.e.}, a combination of planning \& orchestration as well as the YouTube-specific prompt --- strongly outperforms all other setups. 

The comparison to \textit{multi\_agent} reveals that client-specific playbooks provide a crucial lift, even when derived from just two manually completed examples. This offers significant practical value for large-scale migrations, where a small set of "golden" examples can bootstrap quality. However, this domain knowledge only yielded improvements when coupled with the planner and orchestrator (\textit{multi\_agent\_yt\_specific}). For single agents, the additional context ($\sim$3k lines) proved overwhelming, often leading to incomplete migrations where the agent declared success despite implementing only a skeleton of the required logic.

The difference between \textit{multi\_agent} and the single agents shows that the migration quality improves when a multi-agent system is used, at the cost of higher compute, as soon as the total model complexity reaches a certain level. This applies even when no client-specific playbook can be created due to the lack of manual migration examples.

\section{Threats to Validity} \label{section:threats}

We based our evaluation on a representative set of YouTube models, but it’s only a small fraction of all the models that have to be eventually migrated. Experts manually inspected the migrated models and concluded that our metrics are sound. As many of the models are not yet deployed we didn’t account for potentially subtle errors introduced by the system in the migrated models. A more precise measure of the impact and time saving will happen after a large swath of the migrated models is  deployed in production.
We only focused on YouTube models whose size and complexity pose a significant challenge for manual migration. We assume that our conclusions will apply to other families of models, which we have not yet verified beyond a few models used by Google Search where we had the same conclusions.

As noted in Section \ref{section:conclusion}, we will extend our approach to the steps beyond the coding-only part of the migration.

\section{Related Work} \label{section:related}
In the field of automated code transformation there has been a recent emergence of sophisticated multi-agent ecosystems. Our work sits at the intersection of three patterns: task decomposition through orchestration and specialization, domain knowledge distillation, and the transition toward project-scale evaluation.

\subsection{Orchestration and Specialization}
A frequent pattern in recent literature is the use of specialized agent roles to handle long-horizon software engineering tasks. RefAgent \cite{oueslati2025refagentmultiagentllmbasedframework} uses four specialized agents --- planner, refactoring generator, compiler and tester --- to iteratively refine solutions. Similarly, HyperAgent \cite{phan2025hyperagentgeneralistsoftwareengineering} decouples reasoning from environment interaction through a central planner and specialized child agents for navigation and editing. MaintainCoder \cite{wang2025maintaincodermaintainablecodegeneration} breaks engineering tasks into phases such as requirement analysis, design and implementation to improve long-term maintainability of the resulting code. TransAgent \cite{yuan2025semanticalignmentenhancedcodetranslation} uses a system of four agents --- code translator, syntax error fixer, code aligner and semantic error fixer, focusing on different types of error fixing and validation. It relies on an execution alignment strategy for validation. Agentless \cite{xia2024agentlessdemystifyingllmbasedsoftware} uses a three-phase approach to fix or make code changes, where the last phase samples reproduction tests for patch validation.

Our system follows this specialization pattern by separating Planning, Orchestration, and Coding. However, while systems like RefAgent utilize distinct agents for testing and compilation, we consolidate building and testing into the coder's internal ReAct workflow. This leverages Gemini's large context window to maintain full context on the implementation process during test-and-fix cycles.

State management is a key concern when using specialized agents. ChatDev \cite{qian2024chatdevcommunicativeagentssoftware} uses a de-hallucination technique to improve reliability of agent-to-agent communication. Similarly, AgentAsk \cite{lin2025agentaskmultiagentsystemsneed} proposes a reinforcement learning method to reduce the risk of handoff-errors between agents. DynamicCheatSheet \cite{suzgun2025dynamiccheatsheettesttimelearning} curates the context between two consecutive generative calls. In contrast, in our system, agents are tasked to provide sufficiently detailed summaries of their work, which are then passed deterministically to future agent invocations. This showed to be a reliable approach for our use case.

\subsection{Knowledge Distillation}
ACE \cite{zhang2025agenticcontextengineeringevolving} uses a Generator-Reflector-Curator pattern for dynamic context adaptation techniques to produce playbooks, both online and offline. A previous experience report on using AI for code migrations at Google \cite{nikolov2025googleusingaiinternal} specified manually created rules in the playbooks. 

Our strategy for generating client-specific playbooks (see Table~\ref{tab:playbooks}) is a hybrid of those approaches. It mirrors the dynamic context adaptation of ACE but focuses on semi-automated, human-supervised curation of ``golden examples'' to equip agents with client-specific knowledge.

\subsection{Project-Scale Evaluation}
To capture the complexity of project-level code transformations, FreshBrew \cite{may2025freshbrewbenchmarkevaluatingai} uses unit-test based evaluation on repository scale tasks, with mechanisms against cheating (e.g., through agents deleting tests). 
SWE-EVO \cite{thai2026sweevobenchmarkingcodingagents} uses unit-tests to score agents on project-scale evolution tasks, showing that models perform much worse than in benchmarks like SWE-BENCH \cite{jimenez2024swebenchlanguagemodelsresolve}.

While we utilized unit tests to evaluate initial open-source migrations --- hiding tests during the migration to maintain integrity --- we found it fragile, insufficient, and prohibitively expensive to rely on execution-based metrics for complex internal models. Instead, we rely on checklist-based LLM judges. This heuristic approach captures architectural fidelity and completeness where standard unit-test based evaluation fails to scale.

\section{Conclusion and Future Work} \label{section:conclusion}

In this paper, we addressed the significant engineering challenge posed by the large-scale migration of deep learning models from legacy TensorFlow implementations to the modern JAX/Flax ecosystem. The architectural shift from stateful, graph-based paradigms to stateless, functional programming—compounded by the need for strict adherence to internal styles and complex metric reporting—makes manual migration prohibitively expensive for large organizations.

Our primary contribution is an automated multi-agent framework that decomposes this complex transformation task into manageable, verifiable steps through a coordinated hierarchy of planning, orchestration, and coding agents. By leveraging a tiered system of playbooks—ranging from general coding standards to client-specific "golden examples"—we successfully imbued our agents with the domain-specific knowledge required to handle real-world production models. Our evaluation on complex YouTube models demonstrates that this agentic approach achieves a significant speedup over human experts while maintaining high architectural fidelity, as validated by our checklist-based LLM judge.

While these results are promising, the migration process extends beyond code translation. The shift to a stateless paradigm imposes a ``debugging tax'' on developers, often stemming from ``infra-overfitting'' where models rely on legacy framework quirks. Future work will focus on "closing the loop" by developing specialized debugging agents capable of autonomously identifying and fixing training instability or inference latency regressions. We also intend to generalize this framework to other large-scale code transformation tasks, further advancing the "virtuous cycle" where AI-driven systems accelerate the development and modernization of the very machine learning infrastructure they rely upon.

\begin{acks}
This work is the result of a collaboration between the Google Core Developer, CoreML, and YouTube teams. We thank key contributors: Antoine Baudoux, Xevi Miró Bruix, Daniele Codecasa, Madhura Dudhgaonkar, Elian Dumitru, Alex Ivanov, Christopher Milne-O'Grady, Ahmed Omran, Niyati Parameswaran, Ivan Petrychenko, Assaf Raman, Jamie Rogers, Stefan Schnabl, Yurun Shen, Maxim Tabachnyk, Niranjan Tulpule, Amin Vahdat, and Jeff Zhou. We also thank the internal Google reviewers.
We used Gemini 3 Pro to optimize the manuscript for clarity and conciseness, and Nano Banana Pro to generate the system architecture diagrams based on human-provided structural logic and sketches. All AI-assisted content was critically evaluated, edited, and validated by the authors to ensure technical accuracy and original intent.
\end{acks}

\bibliographystyle{ACM-Reference-Format}
\bibliography{references}
\end{document}